\documentclass[aps,prd,twocolumn,nofootinbib,amsmath,amssymb,notitlepage,superscriptaddress]{revtex4-1}
\usepackage{graphicx}
\usepackage{color}
\usepackage{amssymb}
\usepackage{subfigure}
\usepackage{natbib}
\usepackage{color}
\usepackage[colorlinks]{hyperref}
\usepackage{amsmath}
\usepackage{mathtools}

\def\aap{A\&A}                

\begin{document}

\title{Inspiral into Gargantua}

\author{Samuel E. Gralla}
\affiliation{Department of Physics, University of Arizona, Tucson, AZ 85721, USA}
\affiliation{ Center for the Fundamental Laws of Nature, Harvard University, Cambridge, MA 02138, USA }
\author{Scott A. Hughes}
\affiliation{Department of Physics, Massachusetts Institute of Technology, Cambridge, MA 02139, USA}
\affiliation{MIT Kavli Institute for Astrophysics and Space Research, Massachusetts Institute of Technology, Cambridge, MA 02139, USA}
\author{Niels Warburton}
\affiliation{MIT Kavli Institute for Astrophysics and Space Research, Massachusetts Institute of Technology, Cambridge, MA 02139, USA}
\affiliation{School of Mathematical Sciences and Complex \& Adaptive Systems Laboratory, University College Dublin, Belfield, Dublin 4, Ireland}

\begin{abstract}
	We model the inspiral of a compact object into a more massive black hole rotating very near the theoretical maximum.  We find that once the body enters the near-horizon regime the gravitational radiation is characterized by a constant frequency, equal to (twice) the horizon frequency, with an exponentially damped profile. This contrasts with the usual ``chirping'' behavior and, if detected, would constitute a ``smoking gun'' for a near-extremal black hole in nature.
\end{abstract}

\maketitle

\section{Introduction}

General relativity imposes a hard upper limit on how fast a black hole can rotate. For a black hole of mass $M$, the angular momentum $J$ must satisfy
\begin{align}\label{limit}
J \leq G M^2/c,
\end{align}
where $G$ is Newton's constant and $c$ is the speed of light (both hereafter set to unity). Above this value, the event horizon disappears and the spacetime contains a naked singularity. It is impossible to spin up a black hole above this limit with any continuous process featuring reasonable matter \cite{PhysRevLett.57.397}, and there is much evidence in favor of the ``cosmic censorship conjecture'' \cite{1969NCimR...1..252P} that \textit{no} generic initial data can produce a naked singularity.
\begin{figure}
	\includegraphics[width=8.5cm]{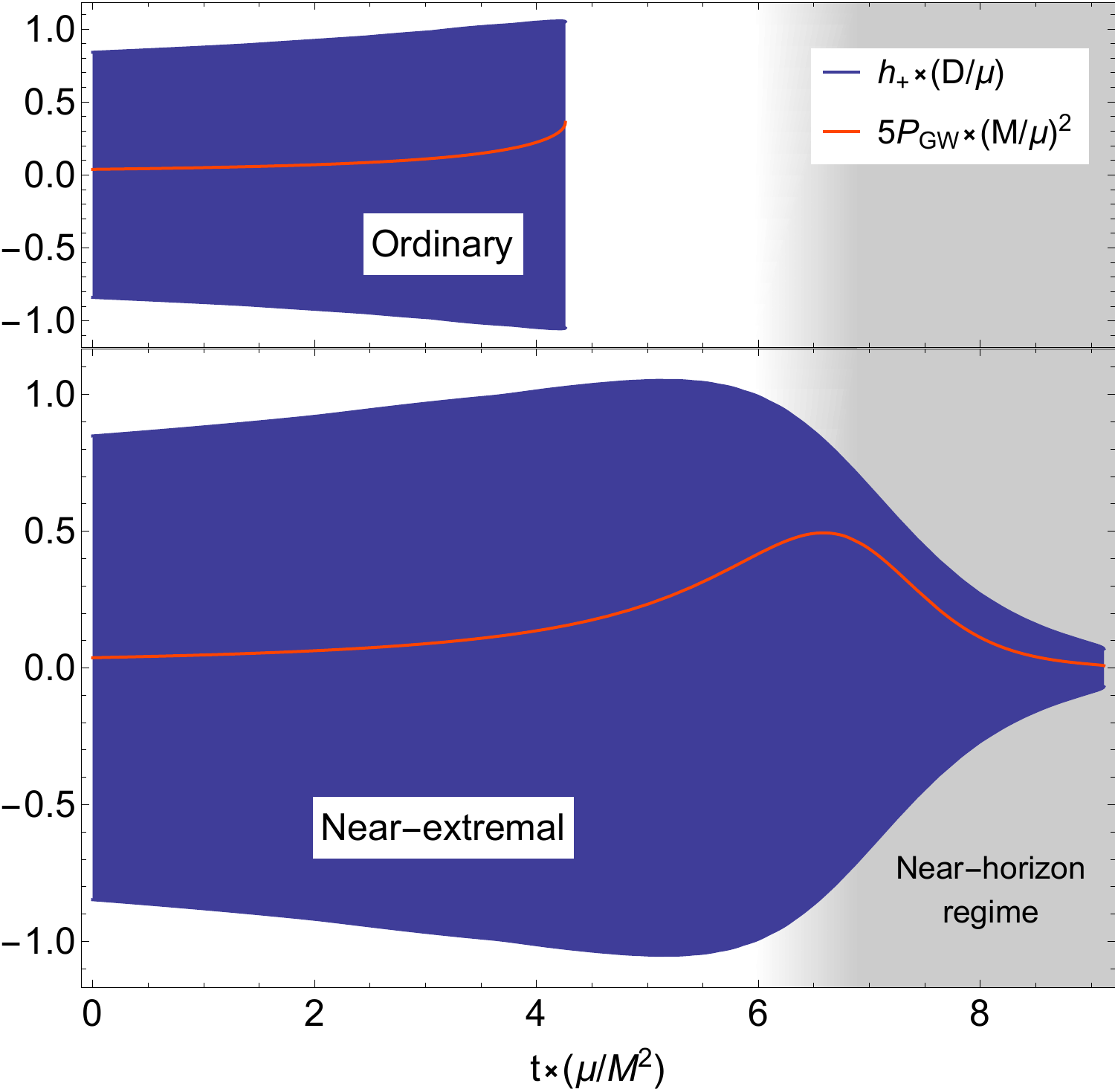}
	\caption{Gravitational waveforms from equatorial, quasi-circular inspiral into ordinary and near-extremal black holes. The black hole spins are $a/M=0.97$ and $a/M=1-10^{-9}$, respectively.  We show the $h_+$ component for a system viewed face-on.  The waveform begins when the particle crosses $r=3.3M$ and ends when the particle reaches the ISCO; we do not model the plunge or ringdown phase of the inspiral in this work.  The individual sinusoidal oscillations of the waveform are too small to see on this scale (where we have assumed a small mass-ratio). We also show (five times) the radiated power, $P_\text{GW}$. The masses of the primary and secondary are denoted by $M$ and $\mu$, respectively, the distance to the binary is $D$.}
	\label{fig:NEK_waveform_example}
\end{figure}

Black holes that saturate the bound \eqref{limit} are known as extremal. More generally, extremal black holes are defined as those with zero Hawking temperature. Extremal black holes play a key role in many theoretical arguments investigating the nature of classical and quantum gravity, such as cosmic censorship \cite{wald1974overspin} and the quantum nature of black hole entropy \cite{strominger-vafa1996}. They have near-horizon regions that possess additional emergent symmetries \cite{bardeen-horowitz1999} and may be governed by a holographic duality \cite{kerrCFT} in the spirit of AdS/CFT \cite{maldacena1998}.  At least in parameter space, they are a hair's breadth from being naked singularities, the existence of which would (in principle) allow experimental study of quantum gravity from a distance.  In light of their basic role in theoretical work, it would be fascinating to discover an extremal black hole in nature.

In this paper we demonstrate a potential means of discovery via a ``smoking gun'': a signal which, if observed, would conclusively establish the presence of a black hole spinning at or extremely near the fundamental limit.  We consider the gravitational radiation from the inspiral of a body into a more massive black hole.  In the non-extremal case, the wave amplitude and frequency increase slowly in time until cutting off rapidly when the compact object reaches the innermost stable circular orbit (ISCO) and plunges into the black hole.  If the black hole is rapidly spinning, however, there is a new, near-horizon phase of the inspiral where the amplitude begins to \textit{decrease} in time and the frequency \textit{saturates} at the horizon frequency -- see Fig.~\ref{fig:NEK_waveform_example}. This can be understood from the fact that the ISCO of a rapidly spinning black hole is close to the horizon and allows access to the near-horizon regime, where the gravitational-wave emission is suppressed because the particle effectively corotates with the black hole. The decay timescale is set by the masses involved, with the total length of the signal set by the black hole spin (diverging in the extremal limit).  The characteristic amplitude decrease is visible in the waveform provided the spin is at least $J\gtrsim0.9999M^2$.

Can black holes of such high spin plausibly exist in nature? From electromagnetic observations there is mounting evidence for a wide distribution in the spins of both stellar mass \cite{Miller:2014aaa} and supermassive black holes \cite{Reynolds:2013rva,brenneman2013}, including measurements consistent, within measurement error, with maximal spin $J=M^2$.  Theoretically, accretion by cold particles can efficiently spin up a black hole to very near the limit \cite{Bardeen:1970}.  However, the inclusion of other effects (such as absorption of hot photons \cite{Thorne:1974ve}) generally limits the spin to more modest values.  The ``Thorne limit'' of $J \lesssim 0.998M^2$ has generally been adopted by the community as a reasonable guess for an astrophysical upper limit.  A method for beating the Thorne limit is discussed in \cite{2011A&A...532A..41S}. 

Our attitude, therefore, is that while near-extremal black holes are perhaps not \textit{expected} to exist in the universe, it is nevertheless highly worthwhile to \textit{search} for them.  The signal we predict is in-band for ground-based detectors \cite{TheLIGOScientific:2014jea,TheVirgo:2014hva} at ``intermediate'' mass ratios and for space-based detectors \cite{Seoane:2013qna} at ``extreme'' mass ratios, with horizon distances similar to ordinary black hole binaries at these mass ratios.  For intermediate mass ratios, one may worry about the validity of our approximation and, with current detectors, about potential confusion with the ringdown of a non-extremal black hole.  We are optimistic on both counts, but further work is required.  For the extreme mass ratios observable by the planned space-based detector eLISA, however, detection would be unambiguous.



Going beyond the classical gravity measurement, the Kerr/CFT conjecture \cite{kerrCFT} holds that near-horizon, near-extremal processes have a dual CFT description.  Aspects of this correspondence have already been tested for bodies orbiting in the near-horizon region \cite{porfyriadis-strominger2014,hadar-porfyriadis-strominger2014,hadar-porfyriadis-strominger2015}.  For the inspiral calculation we perform here, the dual process corresponds to return to equilibrium after a quantum quench \cite{hadar-porfyriadis-strominger2014}.  If the conjecture is correct, observation of near-horizon inspiral constitutes experimental study of strongly coupled quantum field theory.

Other known methods of measuring black hole spin are unlikely to have the precision to discriminate between the Thorne value of $J = 0.998M^2$ and the higher values where the unique features of the extremal case become important.  Thus, to the many exciting possibilities of the new astronomy of gravitational waves \cite{LIGO_detection} we may add one more: the discovery and study of extremal black holes in nature.

The paper is organized as follows.  First we provide an analytical derivation of the basic ``smoking gun'' for quasi-circular inspiral.  We then, in Sec.~\ref{sec:numerical_inspirals}, use numerical methods to explore near-horizon inspiral more generally, showing that the signal is stable to small eccentricity and that its basic features survive the introduction of modest inclination.  In Sec.~\ref{sec:detectability} we consider detectability with current and planned detectors, and Sec.~\ref{sec:gargantua} provides some concluding thoughts.

\section{The Smoking Gun}\label{sec:toke}
We consider a body of mass $\mu \ll M$ (modeled as a point particle) on a prograde\footnote{We do not consider retrograde orbits as they do not enter the near horizon regime during the inspiral phase.} orbit of a Kerr black hole of mass $M$ and angular momentum $J=aM$.  For clarity, we first consider a circular, equatorial orbit, where analytic results are possible.
The particle's energy and angular frequency are given in terms of its Boyer-Lindquist (BL) coordinate radius $r_0$ by  \cite{bardeen-press-teukolsky1972}
\begin{align}
	E 				&= \mu \frac{1-2v_0^2 + \tilde{a}v_0^3}{\sqrt{1-3v_0^2+2\tilde{a}v_0^3}},	\label{eq:E_circ_eq}		\\
	\Omega_\varphi 	&= \frac{M^{1/2}}{r_0^{3/2} + aM^{1/2}},									\label{eq:Omega_phi}	
\end{align}
where $\tilde{a} = a/M$ and $v_0=\sqrt{M/r_0}$.  The ISCO has radius
\begin{align}
	r_\text{ISCO}/M	&= 3 + Z_2 - [(3-Z_1)(3+Z_1+2Z_2)]^{1/2},								\label{eq:r_ISCO}
\end{align}
where $Z_1 = 1 + (1 - \tilde{a}^2)^{1/3}[(1+\tilde{a})^{1/3} + (1-\tilde{a})^{1/3}]$ and $Z_2 = (3\tilde{a}^2 + Z_1^2)^{1/2}$.

\subsection{Near-Horizon Inspiral}

For near-horizon, near-extremal physics it is convenient to introduce dimensionless quantities 
\begin{align}
	\epsilon 	&= \sqrt{1-a^2/M^2},			\\
	x 			&= \frac{r - r_+}{r_+},
\end{align}
where $r$ is the BL coordinate radius and $r_+=M+\sqrt{M^2 - a ^2}$ is the horizon radius. Substituting into Eq.~\eqref{eq:r_ISCO} and expanding, we see that to leading order in $\epsilon$ the ISCO is located at
\begin{align}
x_{\rm ISCO} = 2^{1/3} \epsilon^{2/3},
\end{align}
and hence is in the near-horizon region of a near-extremal black hole.  We will consider a body orbiting outside the ISCO but still in the near-horizon region, i.e.,\footnote{We note that there is plenty of space in the near-horizon region for the particle, in the sense that a body of proper size $R$ subtends a coordinate region $\delta x \sim (R/M)x$.  Thus the point particle approximation is valid as long as the body is small compared to the black hole.}
\begin{align}\label{conditions}
	\epsilon \ll 1, \quad x_\text{ISCO} < x_0 \ll 1.
\end{align}
(More precisely, we let $\epsilon \sim \lambda$ and $x_0 \sim \lambda^{2/3}$ and count orders in $\lambda$.)    Truncating at subleading order, the orbital energy \eqref{eq:E_circ_eq} and azimuthal frequency \eqref{eq:Omega_phi} are
\begin{align}
E 				& = \frac{\mu}{\sqrt{3}}\left[ 1 + \frac{2 x_0}{3}\left( 1 + \frac{\epsilon^2}{x_0^3} \right) \right]		\label{eq:energy} \\
\Omega_\varphi 	& = \frac{1}{2M}\left[ 1 - \frac{3x_0}{4} \right]. \label{eq:Omega}
\end{align}
Note that $1/(2M)$ is the horizon frequency of an extremal Kerr black hole.  In Ref.~\cite{Gralla:2015rpa} it was shown that to leading order such a particle radiates energy in gravitational waves at the rate 
\begin{align}\label{eq:Edot}
	P_{\rm GW} = (C_\infty + C_H) x_0,
\end{align}
where $C_\infty>0$ and $C_H<0$ are dimensionless constants given approximately by\footnote{There are also subdominant oscillations that can be ignored in this analysis. To obtain the numbers in \eqref{eq:Cinf_CH} required keeping up to $\ell=30$ in the sum whose terms are given by Eqs.~(76) and (77) of \cite{Gralla:2015rpa}.}
\begin{align}\label{eq:Cinf_CH}
	C_\infty = 0.987(\mu/M)^2,\quad C_H = -0.133(\mu/M)^2.
\end{align}
These constants give the rate at which energy is radiated to infinity and the horizon, respectively, with the minus sign indicating that energy is being extracted from the black hole.  The energy loss rate translates into an orbital decay rate via $dE/dt=-P_{\rm GW}$. Differentiating \eqref{eq:energy} with respect to $x_0$ and combining with \eqref{eq:Edot} gives a differential equation for the evolution of the orbital radius
\begin{align}\label{eq:dx0_dt_NEK}
	\frac{dx_0}{dt} = -\frac{x_0}{\tau}\left(\frac{1}{1-2\epsilon^2/x_0^3}\right),
\end{align}
where we introduce the timescale
\begin{align}\label{eq:tau}
	\tau \equiv \frac{2}{3\sqrt{3}}\frac{\mu}{(C_\infty + C_H)} = 0.451\mu (M/\mu)^2.
\end{align}
In the above derivation we have retained only leading-order terms in $x_0 \ll 1$.

We have used the instantaneous energy balance relation $dE/dt=-P_{\rm GW}$, which is only valid if the inspiral is evolving adiabatically, i.e., if the orbital timescale $1/\Omega_\varphi$ is much shorter than the inspiral timescale $|x_0/(dx_0/dt)|$.  From \eqref{eq:dx0_dt_NEK} and \eqref{eq:Omega_phi} we see that this happens provided
\begin{align}\label{eq:adiabaticity}
	\textrm{adiabaticity: \ }\frac{\mu}{M} \ll 0.225\left(1-\frac{2\epsilon^2}{x_0^3}\right).
\end{align}
Since $\mu/M$ is small by assumption (and can be at least as small as $10^{-9}$ astrophysically), the inspiral is adiabatic until very close to the ISCO (where the right-hand side vanishes).

We solve Eq.~\eqref{eq:dx0_dt_NEK} exactly below, but for illustration purposes it is useful to consider the case where the particle has entered the near-horizon regime but is still far from the ISCO, i.e., $x_{\rm ISCO} \ll x_0 \ll 1$.  In this case the term $2\epsilon^2/x_0^3=(x_{\rm ISCO}/x_0)^3$ in \eqref{eq:dx0_dt_NEK} is negligible and the equation is trivially solved by
\begin{align}\label{smokey}
	x_0(t) = X_0 e^{-t/\tau}.
\end{align}
where $X_0=x_0(0)$ is the position at the (somewhat arbitrary) point where we declare the beginning of the near-horizon phase of inspiral.  From Eq.~\eqref{eq:Edot} the radiated flux is proportional to the particle's radius and thus the power measured in the gravitational wave detector will drop off exponentially,
\begin{align}\label{eq:Edot_detector}
	P_{\rm detector} \sim e^{-t/\tau}.
\end{align}
Furthermore, from \eqref{eq:Omega} the orbital frequency will increase towards the extremal horizon frequency $\Omega_H=1/(2M)$ as time increases.  The characteristic frequency of the waveform is then twice this frequency (owing to the spin-$2$ nature of gravitation),
\begin{align}\label{eq:Omega_detector}
f_{\rm detector} \sim 2\times\Omega_H/2\pi = \frac{1}{2\pi M}.
\end{align}
Eqs.~\eqref{eq:Edot_detector} and \eqref{eq:Omega_detector} are our basic smoking gun.
\subsection{Time to Plunge}

To estimate the length of time spent in the pre-ISCO inspiral we return to Eq.~\eqref{eq:dx0_dt_NEK}, whose exact solution is
\begin{align}\label{traj}
	x_0(t) = X_0 e^{-t/\tau}g(t),
\end{align}
where we have chosen $x_0(0) = X_0$ and defined
\begin{align}
	g(t) 	&= \exp\left[\frac{k}{3} + \frac{1}{3}W(-ke^{3t/\tau-k})\right],						\label{eq:g(t)}\\
	k 		&= \left(\frac{x_\text{ISCO}}{X_0}\right)^3 = \frac{2\epsilon^2}{X_0^3}. \label{eq:k}
\end{align}
Here $W(x)$ is the Lambert $W$ (or product log) function, defined for $x\geq-1/e$ so that $W(0)=0$ and $W(-1/e)=-1$.  The particle reaches the ISCO when the argument of the Lambert function reaches its limit $-1/e$, which occurs at
\begin{align}\label{tNHI}
	t_\text{NHI} = \frac{\tau}{3}(k - 1 -\log k).
\end{align}
Here NHI is for ``near-horizon inspiral''; $t_{\rm NHI}$ is the length of BL time spent in adiabatic inspiral after the particle arrives at $X_0$, the (somewhat arbitrary) beginning of the near-horizon region.  This is also the length of the signal seen by the detector at infinity.  It should therefore be possible to measure the black hole spin $\epsilon$ from the length of the signal.  Notice that $t_\text{NHI}$ diverges logarithmically as $\epsilon \rightarrow 0$.  This is consistent with the precisely extremal case $\epsilon=0$,\footnote{The flux in the extremal case \cite{porfyriadis-strominger2014} is identical to the flux under the conditions considered here \eqref{conditions}, making the extremal signal identical to what we derive here.  This agreement is highly non-trivial and not well-understood \cite{Gralla:2015rpa}.} where there is no ISCO and the inspiral phase lasts for infinite observer time.

\subsection{Consistency of near-horizon inspiral}

The gravitational waves emitted by the particle during inspiral will change the mass and spin of the black hole.  The change in mass is given by $dM/dt=C_H x_0$ (which is negative---energy is extracted), and the change in angular momentum is given by $dJ/dt=(dM/dt)/\Omega_\varphi $.  (This latter relationship holds for any quasi-circular inspiral.) The associated change in $\epsilon$ is given by
\begin{align}\label{depsilondt}
\frac{d \epsilon}{dt} = \frac{0.102}{M}\frac{\mu^2}{M^2} x_0.
\end{align}
To find the total change over the inspiral, one should plug in the trajectory $x_0(t)$ [Eq.~\eqref{traj}] and integrate from $t=0$ to $t=t_{\rm NHI}$ [Eq.~\eqref{tNHI}].  We can do so analytically if we keep to leading order in $x_{\rm ISCO}/x_0=k^{1/3}$, using $x_0=X_0 e^{-t/\tau}$ and $t_{\rm NHI}=-(\tau/3) \log k$ instead of the exact expressions.  This approximation lengthens the inspiral and makes more of it take place at larger radii where $d\epsilon/dt$ [Eq.~\eqref{depsilondt}] is larger.  Thus the calculation produces an upper bound on the actual change in $\epsilon$, which we find to be
\begin{align}
\Delta \epsilon_{\rm total} < 0.046\frac{\mu}{M}(X_0-2^{1/3} \epsilon^{2/3} ).
\end{align}
For a bound independent of the initial (near-extremal) spin we may drop the term involving $\epsilon$, giving $\Delta \epsilon_{\rm total} < 0.046(\mu/M)X_0$.  In terms of the spin parameter $a/M\approx1-\epsilon^2/2$ this becomes
\begin{align}
\Delta \left[\frac{a}{M}\right]_{\rm total} < 10^{-3} X_0^2 \frac{\mu^2}{M^2}.
\end{align}
The change is quite small even at modest mass ratios.
 
\section{Completing the Picture}\label{sec:numerical_inspirals}

Thus far we have restricted to circular, equatorial inspiral in the near-horizon region, where analytic expressions are available \cite{Gralla:2015rpa}.  To explore near-extremal inspiral more generally we turn to numerics.  In \cite{Gralla:2015rpa} a new code was presented that is capable of working very near extremality.  We also make several improvements to the code of \cite{Throwe:in_prep,O'Sullivan:2014cba} that extend its reach toward extremality.  The numerical values of the flux used in this section are computed using these two codes. Once the flux is known we use the formulae in \cite{Hughes:1999bq,Hughes:2001jr,Drasco:2005kz,Glampedakis:2002ya} to compute the corresponding inspirals and their associated waveforms.

\subsection{Full Quasi-Circular Inspiral}

\begin{figure}[t]
	\includegraphics[width=8.5cm]{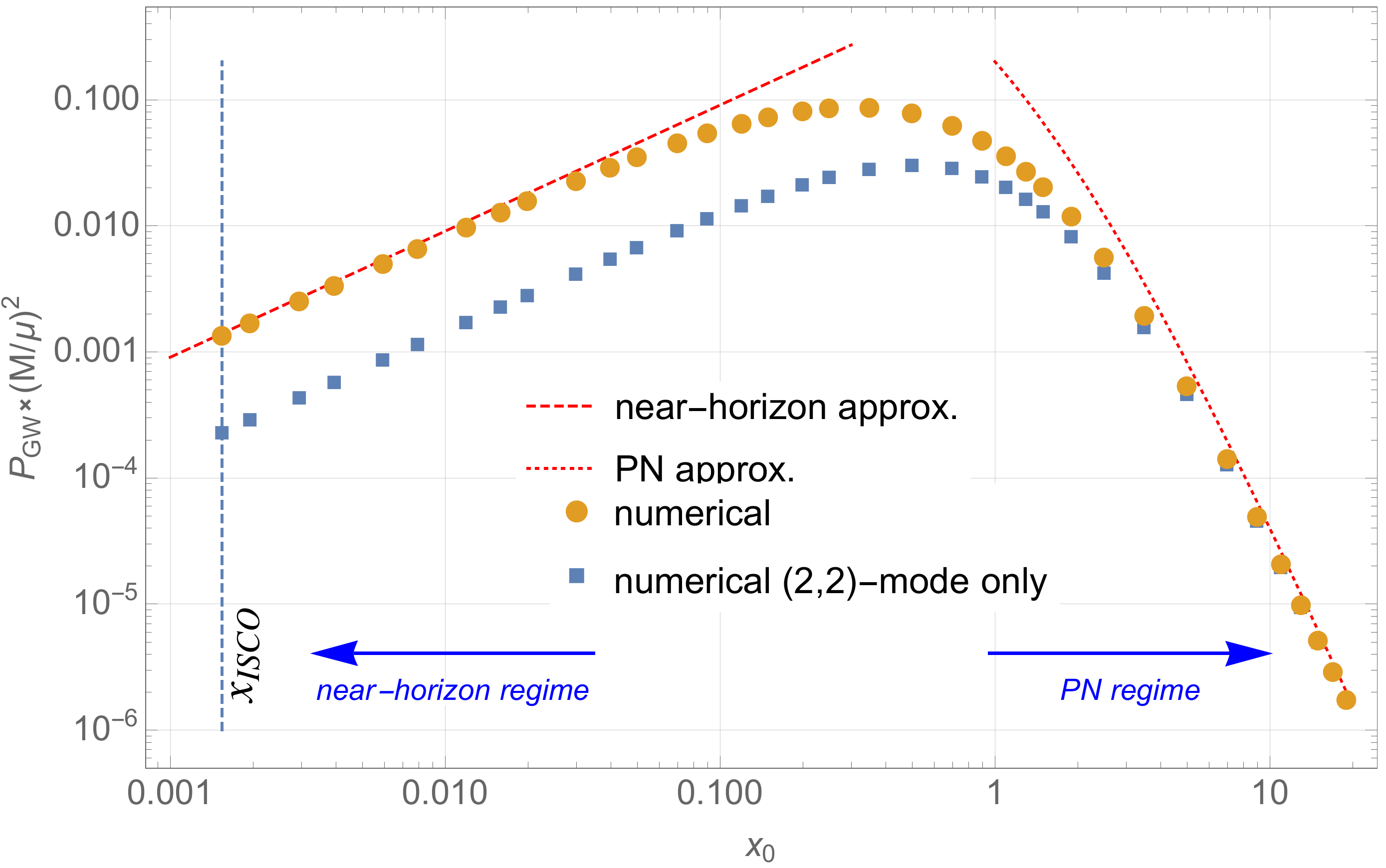}
	\caption{Radiated energy flux from a particle on a prograde, circular, equatorial orbit of a near-extreme Kerr black hole with $a/M=1-10^{-9}$ ($\epsilon=4 \times 10^{-5}$). The radius of the ISCO is marked at $x_{\rm ISCO}\simeq1.6 \times 10^{-3}$.
	}\label{fig:NEK_fluxes}
\end{figure}

\begin{figure*}[t]
\centering
\subfigure[\ Evolution of the orbital radius (solid curve) and frequency (dashed curved). Note the different scales on the left- and right-hand axes.]{\label{fig:left}\includegraphics[width=8.5cm]{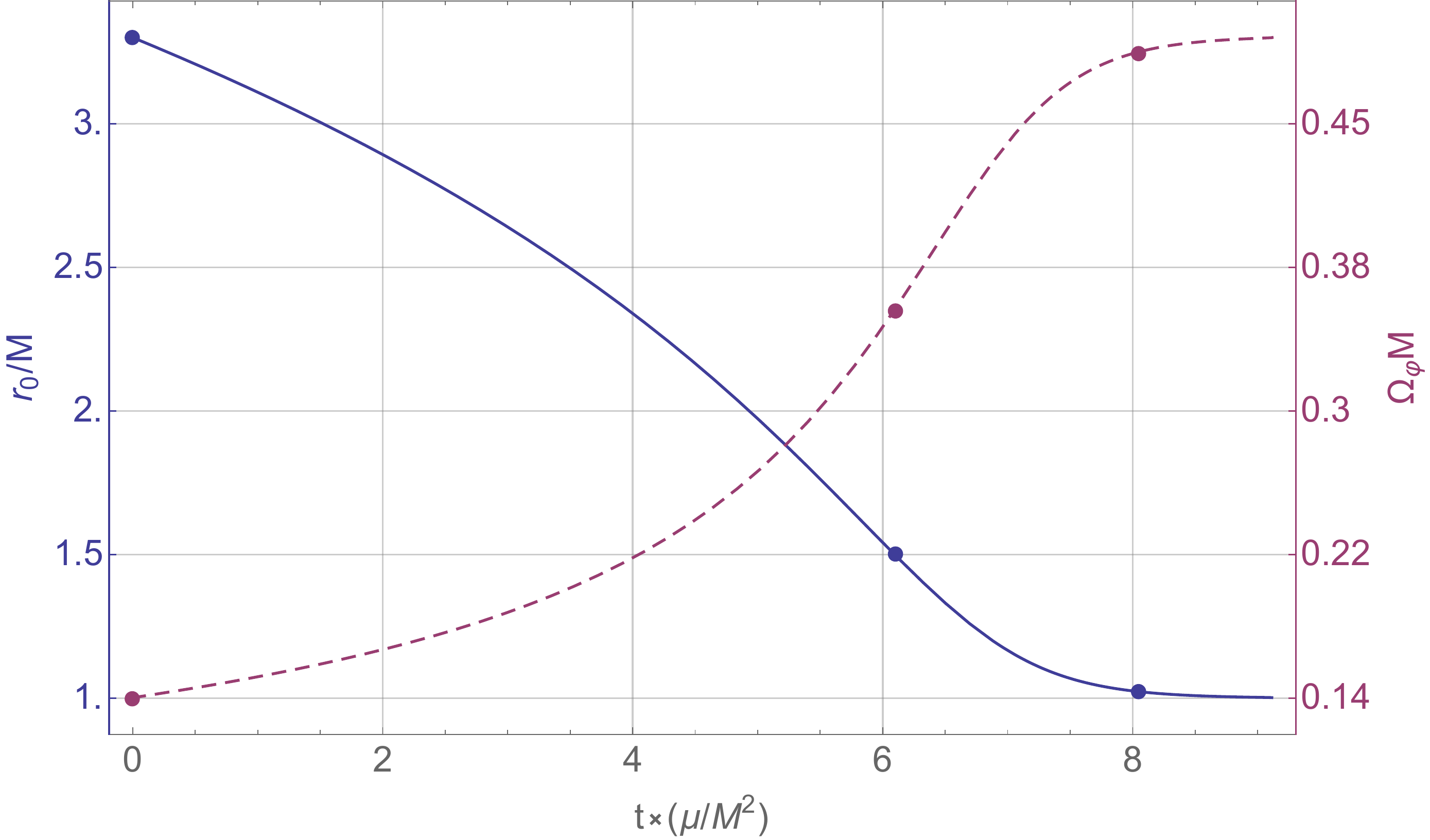}}
\subfigure[\ Edge-on waveforms snapshots. The top, middle and bottom panels occur at the times indicated by the left-most, middle and right-most dots in the left-hand figure.]{\label{fig:right}	\includegraphics[width=8.5cm]{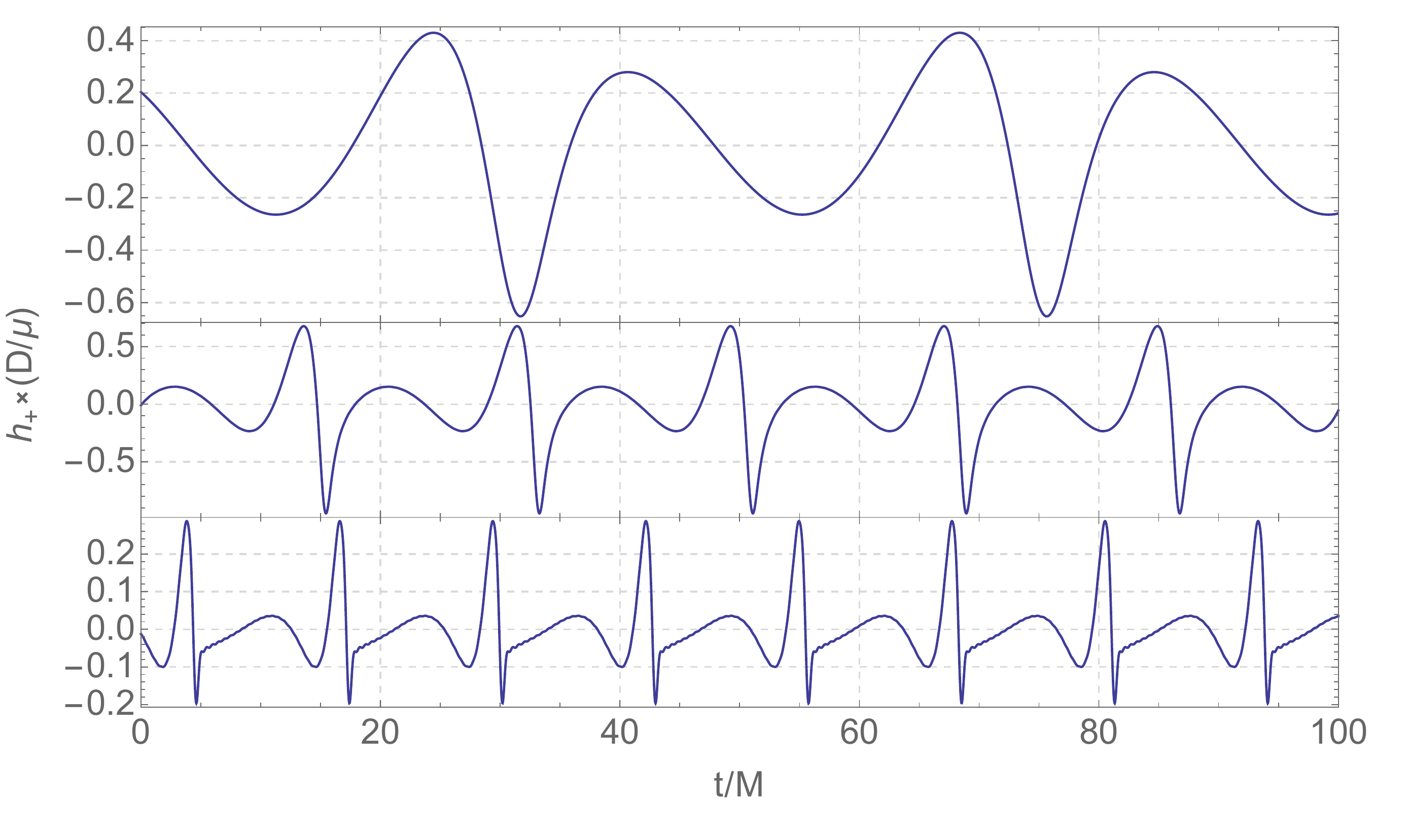}} 
\caption{Quasi-circular, equatorial, inspiral into a near-extreme Kerr black hole with spin parameter $a/M=1-10^{-9}$. The evolution begins at $t=0$ when the particle crosses $r=3.3M$ and ends when the particle reaches the ISCO. }
\label{fig:inspiral}
\end{figure*}

We have numerically computed the flux for stable, circular, equatorial orbits about a near-extreme black hole over the entire range of orbital radii -- see Fig.~\ref{fig:NEK_fluxes} for an example.\footnote{As long as we are outside the ISCO the precise value of near-extremal spin makes little difference to the flux at fixed BL radius.} The flux peaks around $x_0\simeq0.5$ and decays rapidly away from this value.  As a check on our numerical results we compare with analytic approximations near the horizon [Eq.~\eqref{eq:Edot}] and near infinity (using the leading post-Newtonian term $P_{\rm GW} = 32/5(x_0+1)^{-5}$), finding the expected agreement. In Fig.~\ref{fig:NEK_fluxes} we also show the contribution from just the $(l,m)=(2,2)$ mode.  In contrast to the situation at large radii where this mode dominates, in the near-horizon regime it accounts for only $\sim10\%$ of the total flux.  Correspondingly, we require a large number of $\ell$-modes for the total flux to converge in this region (approximately 30 for three-digit accuracy), compared to just a few at larger radii.

We use these fluxes to build a complete adiabatic inspiral, beginning a modest distance away and proceeding through the near-horizon region to the ISCO.  The orbital frequency is seen to monotonically increase, and the orbital radius to monotonically decrease, eventually approaching the horizon values exponentially in time, confirming the analytic prediction -- see Fig.~\ref{fig:left}. The associated waveform when the binary is viewed face on is depicted in Fig.~\ref{fig:NEK_waveform_example}.  Notice that the maximum power is offset from the maximum amplitude, another unique feature that occurs because the frequency continues to rapidly increase even as the amplitude decreases.  When the binary is viewed edge on the waveform exhibits pronounced relativistic beaming in the near-horizon region -- see Fig.~\ref{fig:right}.  This is to be expected as the relevant limit $x_0 \sim \epsilon^{2/3} \rightarrow 0$ is ultra-relativistic, with the circular orbit approaching the null generators of the extremal horizon \cite{jacobson2011}. Similar snapshot waveforms were computed by Detweiler \cite{Detweiler:1978}.

We can use the numerical computations to select an appropriate choice of $X_0$, the boundary of the near-horizon region.  From Fig.~\eqref{fig:NEK_fluxes}, an appropriate choice is $X_0=0.1$, where the numerical expressions begin to diverge from the analytical expressions.  Using Eqs.~\eqref{tNHI} and \eqref{eq:k}, this translates into a bound $\epsilon<0.003$ or $a>0.999995M$ for seeing a full timescale $\tau$ of exponentially decaying waveform.  However, it is clear from Fig.~\ref{fig:inspiral} that the distinctive amplitude decay starts well before the exponential portion sets in, so a more appropriate choice might be $X_0=0.3$, where the flux begins to decrease with decreasing radius.  If we still apply \eqref{tNHI}, this corresponds to the bound $\epsilon<.016$ or $a>0.9999M$ for seeing a full timescale of amplitude decrease.  This is consistent with our numerical experiments, and we have quoted this value in the introduction.

The most likely astrophysical scenario for the formation of a near-horizon, quasi-circular binary is that the orbiting body inspiraled from a much larger distance via gravitational-wave emission.  Some of these waves will be absorbed by the black hole and change its mass and spin.  If there is no other source of angular momentum spinning up the black hole, then this places a bound on how extremal the black hole can be by the time the body reaches the near-horizon region.  Repeating the analysis of \cite{Kesden:2009ds} for our quasi-circular inspiral, we find that an initially extremal black hole will have a spin of $a/M = 1-0.043\mu/M$ by the time the compact object enters the near-horizon regime.  We require $a\ge0.9999M$ in order for the near-horizon waveform to be observed for one timescale, translating  to a restriction $\mu/M < 2\times10^{-3}$ on the mass ratio.  This is no restriction for eLISA sources, but creates some tension for LIGO inspirals. We emphasize, however, that our results for the near-horizon portion are agnostic as to the formation scenario.  We leave it to nature---and observation---to determine whether near-horizon, near-extremal binaries exist at any given mass ratio.

\subsection{More General Inspirals}

Fully generic orbits can be characterized by three orbital parameters: the eccentricity $e$, semi-latus rectum $p$, and inclination $\theta_\text{inc}$. We use the definitions of \cite{Drasco:2005kz}. While we leave the fully generic case for future work, we take the opportunity to point out a number of interesting features of eccentricity and inclination in the near-extreme case.  

\begin{figure}
	\includegraphics[width=8.5cm]{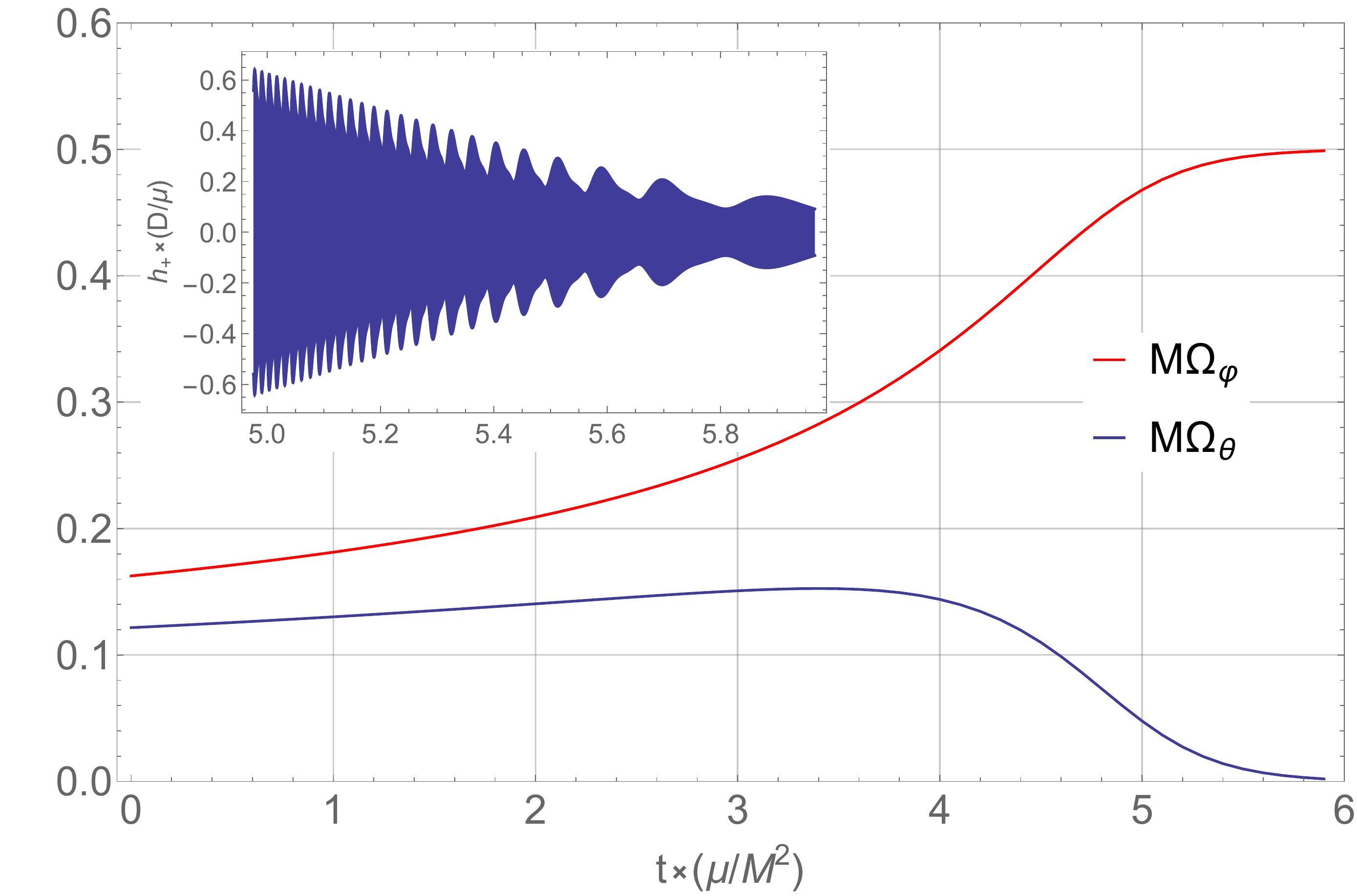}
	\caption{Evolution of the polar and azimuthal orbital frequencies of an inclined (spherical) inspiral into a near-extremal Kerr black hole with $1-a/M=10^{-9}$. The evolution begins at $t=0$ when $r_0=3M$ and $\theta_\text{inc} = 15^\circ$. The inset shows the near-horizon portion of the associated face-on waveform with a decaying envelope modulated by the polar librations of the inspiral. }\label{fig:spherical}
\end{figure}

First, we consider spherical orbits \cite{wilkins1972}, which have a constant coordinate radius $x_0$ and librate in $\theta$ about the equatorial plane up to some maximum value $\theta_{\rm inc}$.  An inspiral which is initially spherical will remain spherical throughout the entire evolution \cite{Hughes:1999bq} and such inspirals have been studied at non-extreme values of spin \cite{Hughes:2001jr}.  In the near-extremal case, there are stable orbits in the near-horizon region provided $\theta_{\rm inc} \lesssim 25^\circ$. In such an inspiral, as the radius of the orbit approaches the horizon, the polar libration frequency $\Omega_\theta$ (defined relative to BL time $t$) approaches zero, while the orbital frequency $\Omega_\varphi$ approaches the horizon frequency.  Fig.~\ref{fig:spherical} shows this behavior in a numerically computed inspiral along with the associated gravitational waveform as an inset.  In the near-horizon regime the waveform exhibits a decaying envelope similar to the equatorial case with the addition of modulations relating the polar libration of the inspiral. As the particle approaches the horizon the polar frequency tends to zero and the associated modulations in the waveform lengthen.  When the initial inclination is small (not shown), the waveform closely matches the corresponding quasi-circular inspiral.

\begin{figure}
	\includegraphics[width=8.5cm]{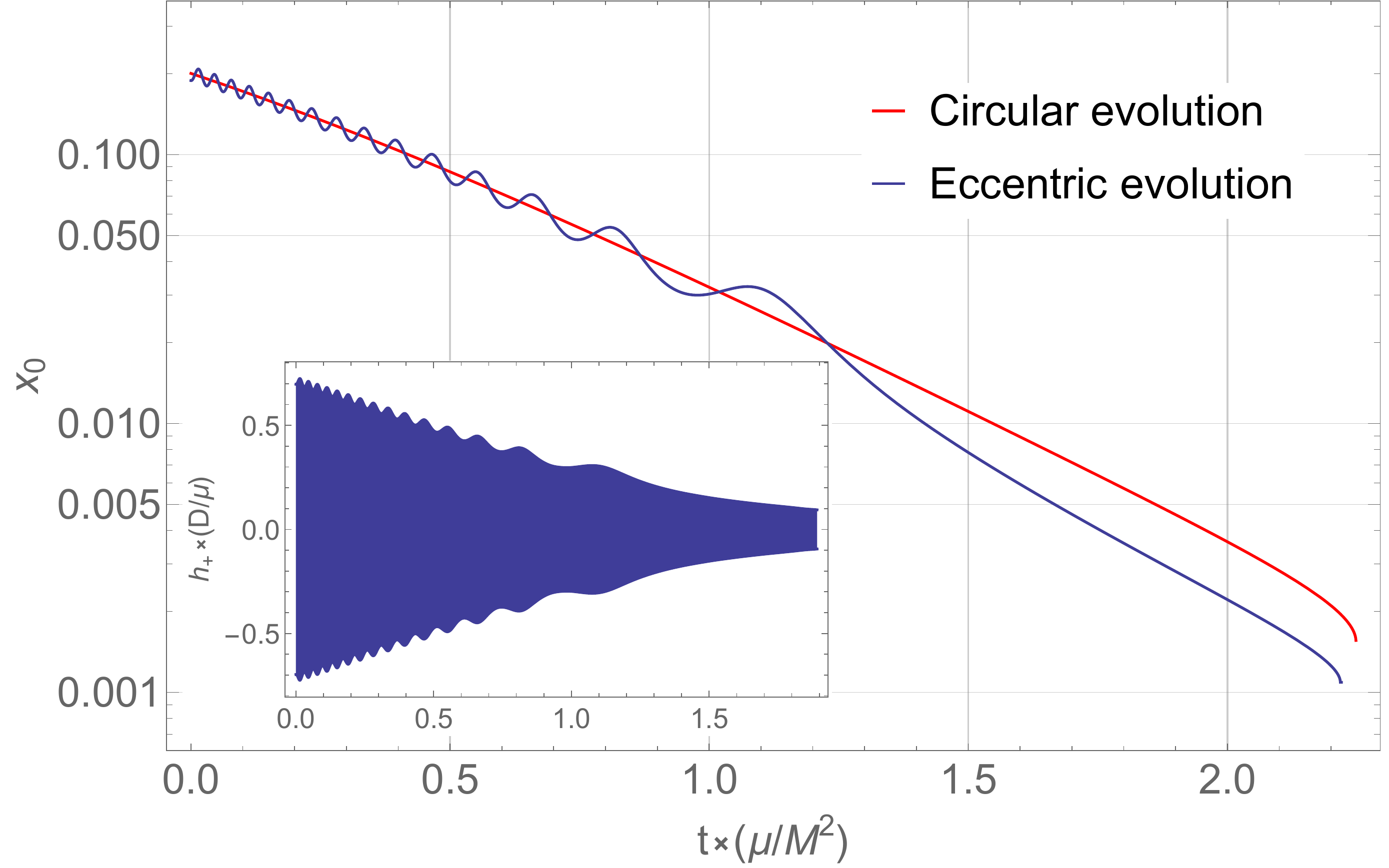}
	\caption{Evolution of a low eccentricity, equatorial inspiral in the near-horizon regime of a Kerr black hole with $a/M=1-10^{-9}$ (blue curve). Initially the orbital eccentricity and semi-latus rectum are $e_0=0.01$ and $p_0=1.2M$, respectively. The eccentric inspiral closely tracks a circular, equatorial inspiral (red curve) that starts with an initial radius of $r_0=p_0$. The inset shows the waveform associated with the eccentric inspiral which has the characteristic exponentially decaying envelope modulated by the radial librations of the inspiral. }\label{fig:ecc}
\end{figure}

Eccentric inspirals present a numerical challenge because of the large number of radial harmonics required for each $\ell,m$ mode in the near-horizon regime.  For this reason we restrict ourselves to exploring low eccentricity inspirals, leaving the more generic case for future work. Our numerical results demonstrate the following.  First, the quasi-circular inspiral is stable to the introduction of small eccentricity for at least the first few timescales.  In particular, the radial motion of the eccentric inspiral closely tracks the orbital radius of a quasi-circular inspiral -- see Fig.~\ref{fig:ecc}.  After a few timescales the eccentricity has typically evolved to become of order the orbital radius (one naive measure of the breakdown of circularity), but the waveform envelope still displays the characteristic amplitude decrease, and is especially similar to the circular case near the end. 

Second, the orbital eccentricity decreases for nearly the entire inspiral.  This is in contrast to the non-extremal case, where the eccentricity decreases for most of the evolution before increasing as the separatrix (the analog of the ISCO) is approached \cite{Cutler:1994pb}.  In our near-extreme inspiral we observe only a tiny uptick very near the separatrix.  This behavior is in line with the predictions and observations of Ref.~\cite{Glampedakis:2002ya}.

Third, we note an unusual feature of highly eccentric inspirals in the near-extremal case: inverted zoom-whirl behavior.  In any inspiral, orbits near the separatrix display a zoom-whirl character \cite{Glampedakis:2002ya}, with each orbital period containing a number of ``whirls'' near an unstable circular orbit and a ``zoom'' out to large radii.  Normally the whirl phase of the waveform is louder, but in the near-extremal case the whirling occurs in the near-horizon region and the amplitude is accordingly suppressed.  The zoom takes the particle out of the near-horizon region.  Thus the zoom and whirl phases are ``inverted'', with the zoom phase being of greater amplitude.

\section{Detectability}\label{sec:detectability}

We now address detectability with ground- and space-based detectors. For the extreme mass-ratio binaries detectable by eLISA, our leading order in the mass ratio calculation provides an excellent approximation to the waveform.  For the intermediate mass-ratios observable from the ground, the reliability of the approximation is less clear.  However, there is reason for optimism in light of recent work showing that that leading-order black hole perturbation theory does surprisingly well even at comparable mass ratios \cite{Tiec:2014lba}.  In what follows we assume the validity of the approximation.

The beginning of an inspiral into a near-extremal black hole, before the body has reached the near-horizon region, is almost identical to the corresponding inspiral into a more modestly spinning black hole (Fig.~\ref{fig:NEK_waveform_example}).  We can therefore use standard detectability estimates for this portion; see, for example, Ref.~\cite{mandel-etal2008} for discussion relevant to intermediate mass ratio inspiral measured by LIGO, and Ref.~\cite{babak-gair-cole2014} for recent discussion relevant to eLISA measurements.  This analysis suggests that Advanced LIGO could detect these sources out to a few hundred megaparsecs, and eLISA could measure events to a redshift $z \sim 0.7$.

Focusing on near-horizon inspiral, we have performed a basic analysis of the signal-to-noise assuming that \textit{only} this portion is seen (App.~\ref{app:detectability}).  We find that the horizon distances are only modestly changed, which is very promising for both detectors.  However, for LIGO, there is the potential to confuse the signal with quasi-normal ringing (QNR) of a more modestly spinning black hole.  QNR is also an exponential decay with frequency set by the horizon frequency (like NHI), but with the time constant set by the mass and spin of the black hole (see, e.g., Ref.\ {\cite{berti-cardoso-starinets2009}} for a review) instead of the two masses of the binary, Eq.~\eqref{eq:tau}.  For LIGO sources, the NHI timescale is comparable to QNR timescales of non-extremal black holes -- see the appendix for details. As such, there is potential to confuse NHI with the ringdown of a more prosaic black hole.

We expect that this concern can be mitigated by taking into account the inspiral prior to the NHI.  The ISCO of a nearly extremal black hole is likely to be at much higher frequency than the ISCO of the black hole one infers from the QNR interpretation; see the appendix for an example.  The incommensurate properties of the pre-NHI inspiral and the NHI interpreted as QNR should signal that one has not correctly interpreted the system's last gravitational waves.  Although further work is necessary to quantify how well LIGO can distinguish a nearly extremal black hole from that is merely rapidly rotating, we are optimistic that this can be done.

At the lower frequencies of a space-based detector like eLISA, there is no confusing NHI with inspiral into a non-extremal black hole.  First, the QNR timescale agrees with the NHI timescale only when the ringing black hole is itself near-extremal,\footnote{The longest QNR timescale is proportional to $M/\epsilon$ near extremality \cite{extremeQNM,extremeQNM2,extremeQNM3}, requiring $\epsilon \sim \mu/M$ to match the NHI timescale \eqref{eq:tau}.} so both interpretations point to a rapidly spinning black hole.  Second, the QNR timescale is not the whole story near extremality: recent results suggest that coherent superposition of modes results in a $1/t$ decay period before the characteristic exponential decay \cite{extremeQNM}.  Such a decay is not present during NHI.  Finally, the NHI decay spends many orbits in the eLISA band---indeed it is possible that a source could be in NHI for the \textit{entire} eLISA lifetime---offering much more promise to distinguish the signals using higher harmonics (e.g., seeing the relativistic beaming of Fig.~\ref{fig:right}).  For other source parameters, we could also see the earlier inspiral and the characteristic transition to near-horizon inspiral.

\section{Gargantua}\label{sec:gargantua}

The title of our article makes reference to Christopher Nolan's science-fiction epic \textit{Interstellar}, which features a black hole named \textit{Gargantua}.  Thorne \cite{thorne2014science} estimated that Gargantua must have a mass of approximately $10^8 M_{\odot}$ and a spin of at least $J \gtrsim (1-10^{-14}) M^2$ in order to enable key pieces of the film's narrative.  This puts Gargantua well within the near-extremal regime studied here.  The associated frequency of a near-horizon inspiral is $10^{-4}$ Hz (see Eq.~\eqref{eq:fH}), which is in-band for eLISA.  If the companion were a sixty solar-mass black hole like one recently observed \cite{LIGO_detection}, the decay timescale would be $12$ years (see Eq.~\eqref{eq:eLISAdecaytime}).  The associated signal would therefore be visible for the entire eLISA lifetime.  If Gargantua is out there, eLISA just might find it.

\section*{Acknowledgements}

We are indebted to Andy Strominger for the inspiration to look for observational signatures of the near-horizon region of a near-extreme black hole.  We thank Sarp Akcay for sharing his {\it Mathematica} code to accurately compute spin-weighted spheroidal harmonics. In addition, we are grateful to Leor Barack, Alex Lupsasca, Achilleas Porfyriadis, Adam Pound, Andy Strominger, and Maarten van de Meent for helpful discussions.  This work was supported in part by NSF grants PHY--1506027 to the University of Arizona, PHY--1205550 to Harvard University, and PHY--1403261 to MIT.  N.W.~gratefully acknowledges support from a Marie Curie International Outgoing Fellowship (PIOF-GA-2012-627781). Many of our computations were performed on the MIT Kavli Institute computing cluster, which is supported by funds from the Kavli Foundation.

\appendix

\section{Detectability of Near-Horizon Inspiral}\label{app:detectability}

In this appendix, we estimate the distance to which a gravitational-wave detector can measure a near-horizon inspiral.  We begin with a standard result for the SNR of a measured gravitational-wave $h(t)$:
\begin{align}
\left(\frac{S}{N}\right)^2 = 2\int_{-\infty}^\infty \frac{|\tilde
  h(f)|^2}{S_h(|f|)}df\;.
\label{eq:SNR1}
\end{align}
The quantity $S_h(f)$ is the 1-sided spectral density of detector noise, and $\tilde h(f)$ is the Fourier transform of the waveform $h(t)$.  Our goal is not to be comprehensive, but rather to provide reliable estimates of detectability, accurate to a factor $\sim 2$.  As such, we will neglect issues related to the detector antenna pattern, and the location and orientation of the binary on the sky (all of which affect our results by factors of order unity).  We will also focus on the simplest case of circular, equatorial inspiral.  Generalizing to more complicated orbits and realistic sky position and orientation is a substantial task that we defer to later analysis.

For circular and equatorial inspiral, the face-on and edge-on cases bound the possibilities, so we will examine them separately.  For face-on, the signal is almost entirely in the $l = |m| = 2$ mode at frequency $f_{\rm gw} = 2 \times \Omega_{\rm H}/2\pi$, with functional form
\begin{align}
h_{\rm FO}(t) = \frac{\mu}{D}e^{-t/\tau}{\cal A}_{\rm FO}\sin(2 \Omega_{\rm H} t)\;.
\label{eq:hFO}
\end{align}
Here, $D$ is the distance to the system, and ${\cal A}_{\rm FO}$ is the amplitude for the face-on case.  Since the signal is monochromatic, the spectral density of noise can be taken out of the integral.  Invoking Parseval's theorem, we can then rewrite the SNR formula (\ref{eq:SNR1}) as
\begin{align}
\left(\frac{S}{N}\right)^2 = \frac{2}{S_h(\Omega_{\rm H}/\pi)}\int_0^T h(t)^2\,dt\;,
\label{eq:SNR2}
\end{align}
where $T$ is the time over which the signal is measured.  Using the waveform (\ref{eq:hFO}) and the fact that $\tau \Omega_{\rm H} \gg 1$, Eq.\ (\ref{eq:SNR2}) yields
\begin{align}
\left(\frac{S}{N}\right)^2 \simeq \frac{\mu^2}{2D^2}\frac{{\cal A}_{\rm FO}^2}{S_h(\Omega_{\rm H}/\pi)}\tau\left(1 - e^{-2T/\tau}\right)\;.
\label{eq:SNR3}
\end{align}
The leading correction to Eq.\ (\ref{eq:SNR3}) is $O[(\tau\Omega_{\rm H})^{-1}]$.

For edge-on, a large number of modes make a significant contribution to the signal the detector measures.  The signal is dominated by modes with $l = |m|$, which allows us to write the signal as a sum of exponentially decaying sinusoids:
\begin{align}
h_{\rm EO}(t) = \frac{\mu}{D}e^{-t/\tau}\sum_{m = 2}^\infty {\cal A}^m_{\rm EO}\sin(m \Omega_{\rm H} t)\;.
\end{align}
Using this form as well as $\tau\Omega_{\rm H} \gg 1$, the SNR we find in this case is
\begin{align}
\left(\frac{S}{N}\right)^2 \simeq \frac{\mu^2}{2D^2}\tau\left(1 - e^{-2T/\tau}\right)\sum_{m = 2}^\infty
\frac{({\cal A}^m_{\rm EO})^2}{S_h(m\Omega_{\rm H}/2\pi)}\;.
\label{eq:SNR4}
\end{align}
Systems will generally lie somewhere between these two extremes, which bound the possible range.

To see what range this produces, consider measurements at threshold SNR $\rho_{\rm th}$, and use Eq.\ (\ref{eq:tau}) to rewrite $\tau$ in terms of $M$ and $\mu$.  For the face-on case, we find
\begin{align}
D^{\rm max}_{\rm FO} = \frac{M{\cal A}_{\rm FO}}{\rho_{\rm th}}
\sqrt{\frac{0.225\mu(1 - e^{-2T/\tau})}{S_h(\Omega_{\rm H}/\pi)}}\;.
\end{align}
For edge-on, we have
\begin{align}
D^{\rm max}_{\rm EO} = \frac{M}{\rho_{\rm th}}
\sqrt{0.225\mu(1 - e^{-2T/\tau})}\sqrt{\sum_{m = 2}^\infty\frac{({\cal A}^m_{\rm EO})^2}{S_h(m\Omega_{\rm H}/2\pi)}}\;.
\end{align}
To parameterize the wave's amplitude, we compute the gravitational waveform as the small body enters the near-horizon regime ($x_0 \simeq 0.3$, see Fig.\ {\ref{fig:NEK_fluxes}}), and read off
\begin{eqnarray}
{\cal A}_{\rm FO} &\simeq& 0.8\;,
\nonumber\\
{\cal A}^2_{\rm EO} &\simeq& 0.23\;, 
\nonumber\\
{\cal A}^3_{\rm EO} &\simeq& 0.19\;, 
\nonumber\\
{\cal A}^4_{\rm EO} &\simeq& 0.145\;, 
\nonumber\\
{\cal A}^5_{\rm EO} &\simeq& 0.115\;. 
\label{eq:amplitudes}
\end{eqnarray}
We have computed the edge-on amplitudes up to $A^{10}_{\rm EO}$, but those beyond ${\cal A}^5_{\rm EO}$ only change the distances we infer by $\sim 5\%$.  Note that these amplitudes continue to decrease as $m$ increases.

\subsection{Detectability by eLISA}

Let us now consider plausible figures for a near-extreme inspiral observed by eLISA.  Our signal appears at harmonics of
\begin{align}
f_{\rm H} = \frac{\Omega_{\rm H}}{2\pi} = \frac{1}{4\pi M} = 1.6\times 10^{-3}\,{\rm Hz}\left(\frac{10^7\,M_\odot}{M}\right)\;.
\label{eq:fH}
\end{align}
The eLISA sensitivity is expected to be approximately flat in its band of peak sensitivity, from about $3 \times 10^{-3}\,{\rm Hz} \lesssim f \lesssim 10^{-1}\,{\rm Hz}$.  The strain spectral density at these frequencies is $S_h(f) \simeq 4 \times 10^{-40}\,{\rm Hz}^{-1}$ (Fig.\ 12 of Ref.\ {\cite{Seoane:2013qna}}, noting that the vertical axis is the square root of $S_h$).  Most observable harmonics of $f_{\rm H}$ will lie in this nearly flat band for $M \sim 10^7\,M_\odot$.

Although detailed analysis will be needed to determine an appropriate threshold SNR, for a monochromatic signal lasting $\sim 10^5$ cycles, it is likely to be of order 15.  Let us consider a small body of $\mu = 10\,M_\odot$ spiraling into a black hole of $M = 10^7\,M_\odot$.  For this case, $\tau = 0.451 M^2/\mu \simeq 0.7$ years.  Using the amplitudes (\ref{eq:amplitudes}), we find
\begin{eqnarray}
D^{\rm max}_{\rm FO} &\simeq& 4.2\,{\rm Gpc}
\left(\frac{15}{\rho_{\rm th}}\right)
d(\mu, M, T)\;,
\\
D^{\rm max}_{\rm EO} &\simeq& 2.0\,{\rm Gpc}
\left(\frac{15}{\rho_{\rm th}}\right)
d(\mu, M, T)\;,
\end{eqnarray}
with
\begin{equation}
d(\mu, M, T) = 
\left(\frac{M}{10^7\,M_\odot}\right)
\left(\frac{\mu}{10\,M_\odot}\right)^{1/2}
\sqrt{1 - e^{-2T/\tau}}\;.
\end{equation}
These figures suggest that the range of eLISA for nearly extremal inspiral are comparable to the range that has been found for ``ordinary'' extreme mass-ratio inspirals {\cite{Seoane:2013qna}}.

\subsection{Detectability by LIGO}

By Eq.\ (\ref{eq:fH}), we have $f_{\rm H} = 50\,{\rm Hz}$ if $M = 320\,M_\odot$.  At this mass, the $m = 2$ harmonic radiates very close to the peak sensitivity of Advanced LIGO, where the spectral density of noise in its final configuration is expected to reach $S_h \simeq 4\times10^{-46}\,{\rm Hz}^{-1}$.  
Repeating\footnote{Note that Eqs.\ (\ref{eq:SNR3}) and (\ref{eq:SNR4}) strictly speaking do not apply in the LIGO case, since $\tau \Omega_{\rm H}$ is not large.  However, we find empirically that the error one makes using these formulas is percent level or smaller.} our analysis 
using LIGO noise levels, we find
\begin{eqnarray}
D^{\rm max}_{\rm FO} &\simeq& 200\,{\rm Mpc}
\left(\frac{10}{\rho_{\rm th}}\right)
\left(\frac{M}{320\,M_\odot}\right)
\left(\frac{\mu}{10\,M_\odot}\right)^{1/2}\;,
\label{eq:ligoFO}\\
D^{\rm max}_{\rm EO} &\simeq& 100\,{\rm Mpc}
\left(\frac{10}{\rho_{\rm th}}\right)
\left(\frac{M}{320\,M_\odot}\right)
\left(\frac{\mu}{10\,M_\odot}\right)^{1/2}\;.
\label{eq:ligoEO}
\end{eqnarray}
(Note that we use a lower threshold SNR here, since the number of cycles in band is smaller --- the signal is not spread out over as many cycles as in the eLISA case.  Note also that the scaling with $M$ is quite rough, and only approximately accurate over the range $200\,M_\odot \lesssim M \lesssim 500\,M_\odot$.  Outside of this range, the signal moves to frequencies at which the noise deviates significantly from our fiducial value.)  The values (\ref{eq:ligoFO}) and (\ref{eq:ligoEO}) suggest that it might be possible to see near-extremal black hole physics with LIGO, given nearly extremal black holes of a few hundred solar masses.

\subsection{Distinguishing NHI from QNR}

The NHI waveform will unfortunately look very similar to quasi-normal ringing: both are exponentially decaying sinusoids.  As such, one should ask what the consequences are of confusing NHI for QNR.  For the case of detection by eLISA, we find that this confusion is essentially harmless: even if we mistake NHI for QNR, we will conclude that we have observed processes involving a nearly extremally rotating black hole.  For LIGO, this confusion is not harmless, and would lead us to conclude that we have observed inspiral into a non-extremal black hole.

To understand the consequences of these two varieties of confusion, let us examine the exponential decay time for NHI:
\begin{eqnarray}
\tau &=& 0.451\,M^2/\mu
\nonumber\\
&=& 0.023\,{\rm sec} \left(\frac{M}{320\,M_\odot}\right)^2
\left(\frac{10\,M_\odot}{\mu}\right)\;,
\label{eq:LIGOdecaytime}\\
&=& 0.71\,{\rm year} \left(\frac{M}{10^7\,M_\odot}\right)^2
\left(\frac{10\,M_\odot}{\mu}\right)\;.
\label{eq:eLISAdecaytime}
\end{eqnarray}
In Eq.\ (\ref{eq:LIGOdecaytime}), we use the fiducial parameters we selected for our LIGO distance estimation; in Eq.\ (\ref{eq:eLISAdecaytime}), we use our fiducial eLISA parameters.

For our LIGO parameters, this signal decays very quickly.  The system only radiates
\begin{eqnarray}
N_\tau &=& 2f_{\rm H}\tau
\nonumber\\
&\simeq& 2.3\,\left(\frac{M}{320\,M_\odot}\right)
\left(\frac{10\,M_\odot}{\mu}\right)
\end{eqnarray}
cycles in the $m = 2$ mode before its amplitude has fallen by a factor of $1/e$.  Such a signal would be indistinguishable from, and likely be confused with, the quasi-normal ringing of a much more slowly spinning black hole.  Using the fits given in Ref.\ {\cite{berti-cardoso-will2006}}, we find that our fiducial LIGO NHI would be indistinguishable from ringdown for a black hole with mass $M = 240\,M_\odot$, spin $a/M = 0.95$.  We note, however, that the inspiral preceding the NHI waveform is likely to have properties inconsistent with such a relatively slowly spinning black hole.  For example, for a $320\,M_\odot$ black hole, the ISCO corresponds to an $m = 2$ mode of $f = 55$ Hz for $a/M = 0.95$, significantly smaller than the $f = 100$ Hz value found in the near-extremal limit.  Although further analysis is necessary, it seems likely that the incommensurate characteristics of the QNR interpretation of the NHI waves with the pre-NHI inspiral may allow us to probe near-extremal black holes with LIGO.

The NHI signal is quite long-lived in the eLISA band.  For our fiducial parameters, the signal radiates
\begin{eqnarray}
N_\tau \simeq 72,000\,\left(\frac{M}{10^7\,M_\odot}\right)
\left(\frac{10\,M_\odot}{\mu}\right)
\end{eqnarray}
cycles in the $m = 2$ mode before its amplitude falls by $1/e$.  Even if one attempted to interpret this signal as quasi-normal ringing, its very slow falloff would lead one to conclude that the large black hole's spin is nearly extremal.  For example, if one uses the fits given in {\cite{berti-cardoso-will2006}}, one finds that our fiducial eLISA NHI signal looks like ringdown for a black hole with spin $a/M \simeq 1 - 10^{-10}$.  In other words, whether we interpret this signal as near horizon inspiral or as quasi-normal ringing, we conclude that the black hole is nearly extremal.  

\bibliographystyle{apsrev4-1}
\bibliography{gargantua}

\end{document}